# Title: Experimental Realization of Deep Subwavelength Confinement in Dielectric Optical Resonators

**Authors:** S. Hu[1], M. Khater[2], R. Salas-Montiel[3], E. Kratschmer[2], S. Engelmann[2], W. M. J. Green[2], S. M. Weiss[1,4]*

**Affiliations:**

[1]Department of Physics and Astronomy, Vanderbilt University, Nashville, TN 37235, USA.

[2]IBM T J Watson Center, 1101 Kitchawan Road, Yorktown Heights, New York 10598, USA.

[3]Laboratoire de Nanotechnologie et d'Instrumentation Optique, Institut Charles Delaunay CNRS-UMR 6281, Université de Technologie de Troyes, Troyes 10004, France.

[4]Department of Electrical Engineering and Computer Science, Vanderbilt University, Nashville, Tennessee 37235, USA.

*Correspondence to:  sharon.weiss@vanderbilt.edu

**One Sentence Summary:** Dielectric cavities support record low mode volumes by incorporating subwavelength features into photonic crystal unit cells.

**Abstract**: The ability to highly localize light with strong electric field enhancement is critical for enabling higher efficiency solar cells, light sources, and modulators. While deep subwavelength modes can be realized with plasmonic resonators, large losses in these metal structures preclude most practical applications. We developed an alternative approach to achieving subwavelength confinement that is not accompanied by inhibitive losses. We experimentally demonstrate a dielectric bowtie photonic crystal structure that supports mode volumes commensurate with plasmonic elements and quality factors that reveal ultra-low losses. Our approach opens the door to the extremely strong light-matter interaction regime with simultaneously both ultra-low mode volume and ultra-high quality factor that has remained elusive in optical resonators.

**Main Text**:

Light-matter interaction in an optical resonator is enhanced through two confinement mechanisms: (i) temporal confinement, which is the photon cavity lifetime and is characterized by the quality factor (Q) and (ii) spatial confinement, which is the ability to focus light into a tightly confined space and is characterized by the modal volume ($V_m$). Simultaneously achieving high confinement in both categories has been a long-time pursuit in nanophotonics research and holds the promise for revolutionary advances in generating, modulating, and detecting light, including higher efficiency light sources (*1-4*) and solar cells (*5-8*), as well as faster and lower power consumption optical switches and modulators (*9-13*). Plasmonic and metal-based metamaterial structures are capable of concentrating light into deep-subwavelength volumes (i.e., mode volume = $V_m \sim 10^{-3}$ ($\lambda/n_{air})^3$) by accessing a surface plasmon resonance (*6, 14, 15*). However, resistive heating losses lead to poor temporal confinement (i.e., quality factor = Q ~ 10) and prohibit the realization of practical devices that require propagation of energy (*16*). Recent work replacing metals with all-dielectric materials has led to encouraging progress for enhanced spatial light localization through scattering in high index dielectric nanoparticles, but the lack of an intrinsic confinement mechanism within these dielectric structures has prevented light concentration on par with plasmonics. Furthermore, the scattering mechanism in these subwavelength dielectric nanoparticles is incapable of providing temporal confinement (*17-19*).

Historically, low-loss dielectric structures, such as interferometers and ring resonators, have been the building blocks of photonic technologies, but they are diffraction-limited and therefore unable to focus light below $\lambda/2n_d$ where $n_d$ is the refractive index of the dielectric material in which the optical mode is confined. Photonic crystals have provided the best confinement in lossless dielectric materials to date (*20*). Typical photonic crystals utilize a simple unit cell geometry – circles (**Fig. 1, A and B**) or rectangles (*21-23*) – and have a $V_m \sim 1$ $(\lambda/n_d)^3$. Slotted photonic crystal cavities can further squeeze light into a nanoscale low index region by designing an abrupt index discontinuity along the electric field polarization direction (**Fig. 1, C and D**). Slotted photonic crystal cavities reduce the $V_m$ to $\sim 0.01$ $(\lambda/n_{air})^3$, almost two orders of magnitude lower than traditional photonic crystal cavities (*24, 25*). However, it is difficult to achieve deeper subwavelength confinement via a slot configuration alone. Additionally, since slotted photonic structures inherently confine cavity modes within a low index region, they are not suitable for applications requiring strong light-matter interaction in high index materials, such as silicon or many highly nonlinear optical materials.

While the spatial localization of photons typically occurs due to a single physical mechanism, such as total internal reflection in waveguides and photonic bandgap confinement in photonic crystals, it is apparent from previous work that a second level of spatial localization is possible (*6, 15, 24-26*). In the case of slotted photonic crystal waveguides, light is first confined in the dielectric mode by the photonic bandgap effect such that light is spatially localized in the dielectric region between the lattice holes. Then, introduction of an air slot that cuts through the dielectric region further localizes the light within this air slot due to electromagnetic boundary conditions. Our recent theoretical study suggests that this two-step light confinement effect can be best exploited to achieve low mode volume by using subwavelength modifications of the photonic crystal lattice holes rather than the dielectric region between lattice holes (*26*). In that study, light is first confined in the air mode by the photonic bandgap effect such that light is spatially localized inside the lattice air holes. Then a bowtie-shaped subwavelength dielectric inclusion added to the lattice holes enables redistribution of the optical mode to the tips of the bowtie as a result of electromagnetic boundary conditions. Importantly, the two-step light localization process in photonic crystals preserves the high-Q nature of the photonic crystal cavity. In this report, we introduce design improvements to bowtie-shaped photonic crystal unit cells that enable additional mode confinement in the out-of-plane direction (**Fig. 1, E to G**) and we experimentally demonstrate a bowtie photonic crystal cavity with a loaded Q-factor on the order of $10^5$ and record deep subwavelength mode confinement in silicon ($V_m \sim 10^{-3}$ $(\lambda/n_{Si})^3$) that is verified by optical near-field measurements. Such extreme light concentration is on par with plasmonic resonators yet the low-loss dielectric material allows a concurrent ultra-high Q-factor (*26*).

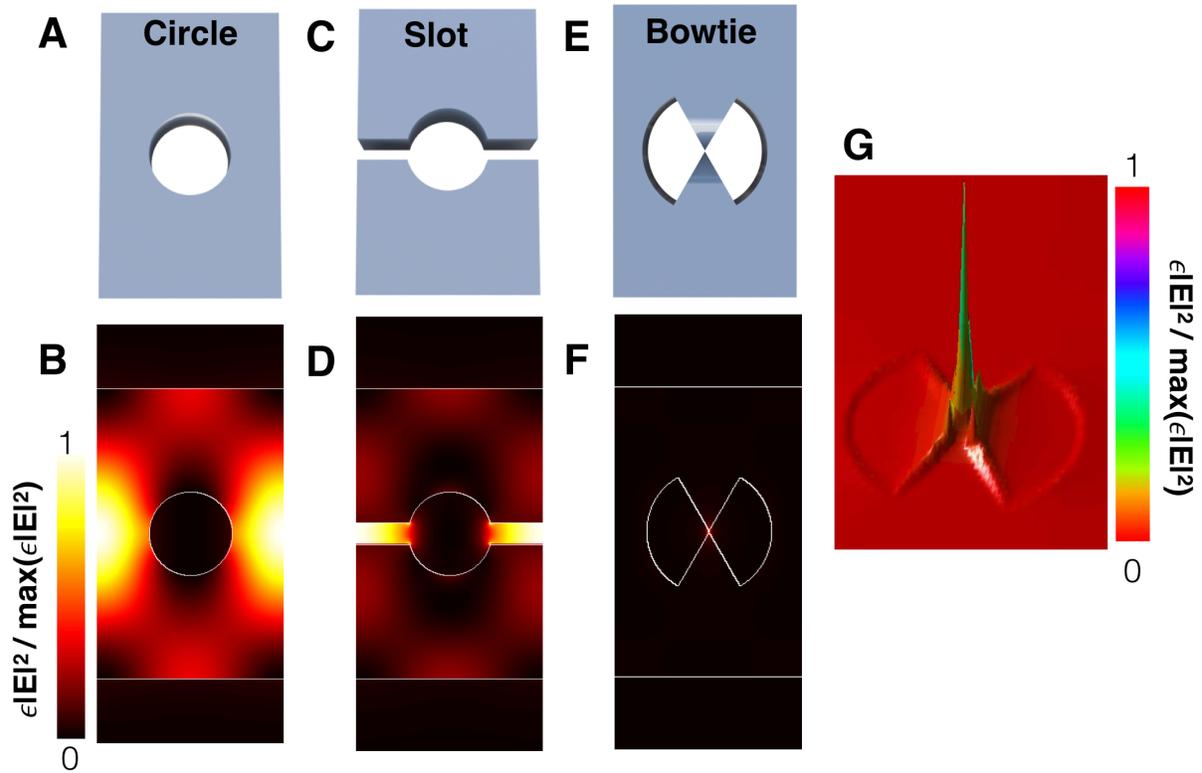

**Fig. 1**. **Comparison of light concentration in different photonic crystal unit cells.** (**A-B**) Traditional circular unit cell of a photonic crystal and its electric energy profile at the dielectric mode band edge. (**C-D**) Slotted photonic crystal unit cell and its mode profile at the dielectric band edge. (**E-F**) Bowtie photonic crystal unit cell and its electric energy profile at the air band edge. (**G**) 3D profile of the optical mode in the bowtie unit cell showing the optical energy distribution. All color maps are scaled according to the minimum and maximum electrical energy density values of each individual unit cell.

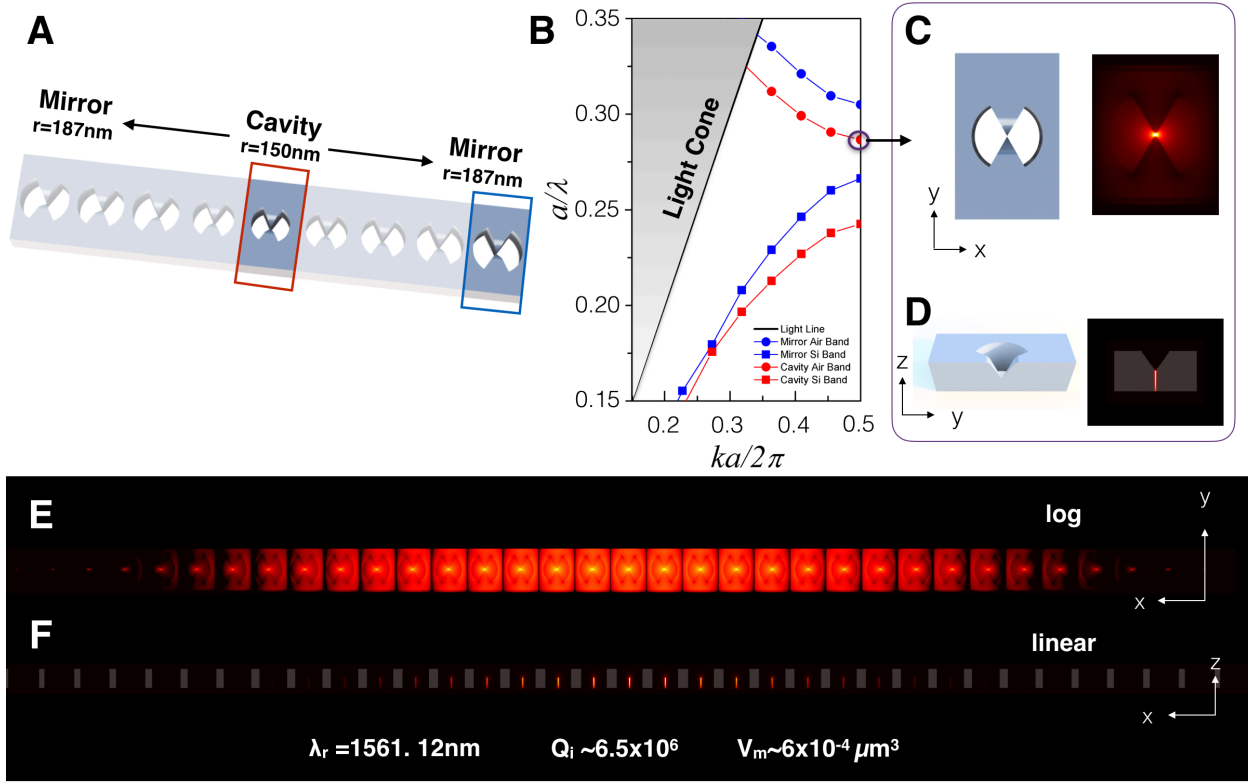

**Fig. 2. Design of silicon photonic crystal using a bowtie shaped unit cell.** (**A**) The cavity is formed with a center unit cell of 150 nm radius and mirror unit cells of 187 nm radii on both sides of the cavity. The radius is gradually tapered from the center to the mirror segments. The photonic crystal lattice spacing is $a$=450 nm and the width of the waveguide is 700 nm. The structure is designed with a 220 nm silicon device layer and 2 μm thick buried oxide layer. (**B**) Optical band structures of the cavity unit cell (red curve) and mirror unit cell (blue curve). (**C**) Top view ($xy$ plane) and (**D**) cross-section view ($yz$ plane) schematics and associated air band edge electrical energy profile in the center unit cell. (**E**) Log plot of the photonic crystal cavity resonance electric energy profile in the $xy$ plane at $z = 0$. (**F**) Linear plot of the photonic crystal cavity in the $xz$ plane at $y = 0$.

In order to achieve improved out-of-plane modal confinement in the bowtie unit cell, a thickness modulation is designed in the bowtie tip region to form a v-groove cross-sectional profile that is experimentally realizable. A silicon photonic crystal cavity is designed using this bowtie-shaped unit cell with the v-groove by slowly varying the radius of the unit cell, as shown in **Fig. 2A**. Since the dielectric bowtie element is inside the envelope of a traditional air hole, the photonic crystal is designed to confine the air mode such that light is localized into the air hole region and then further localized to the tips of the dielectric bowtie within the air hole region. The band diagram in **Fig. 2B** shows that the air mode of the bowtie photonic crystal cavity unit cell lies within the mode gap of the mirror unit cells, providing the requisite confinement for the cavity mode. The wavelength of the cavity air mode is approximately 1570 nm at the band edge ($k_x = 0.5(2\pi/a)$). As is the case for all 1D photonic crystal cavities, the Q-factor of the bowtie photonic crystal cavity is governed in large part by the band gap tapering from the cavity to mirror unit cells. We choose to transition between the center and mirror unit cells in a quadratic fashion, similar to the approach followed in other high Q photonic crystal designs (*21-23*).

**Figure 2C** and **D** show the top and side view profiles of the optical mode in the center cavity unit cell, and the electric energy distribution across the bowtie photonic crystal cavity is shown in **Fig. 2, E and F, and fig. S2A**. In this simulation, there are 20 tapering unit cells between the central cavity unit cell and the 10 mirror unit cells on each side of the cavity; not all unit cells are shown in the figures. The mode is highly confined between the bowtie tips in the central cavity unit cell (**fig. S2, B and C**) and decays gradually into the mirror segments, giving a Gaussian-shaped electric energy profile that minimizes radiation losses (**fig. S2D**). At the resonance wavelength of $\lambda = 1561.12$ nm, the simulated Q-factor is $6.55 \times 10^6$. The $V_m$ is calculated to be $6.09 \times 10^{-4}$ μm$^3$ using **Equation 1** where E is the electric field and ε is the permittivity.

$$V_m = \frac{\int \varepsilon |E|^2 \, dV}{\text{Max}(\varepsilon |E|^2)} \qquad (1)$$

Given the small dimensions of the bowtie tip, the mode extends partially in air and partially in silicon. Hence, the normalized mode volume should fall between that of the mode volume normalized to the wavelength in silicon – $V_m = 6.7 \times 10^{-3}$ $(\lambda/n_{Si})^3$ – and the mode volume normalized to the wavelength in air – $V_m = 1.6 \times 10^{-4}$ $(\lambda/n_{air})^3$.

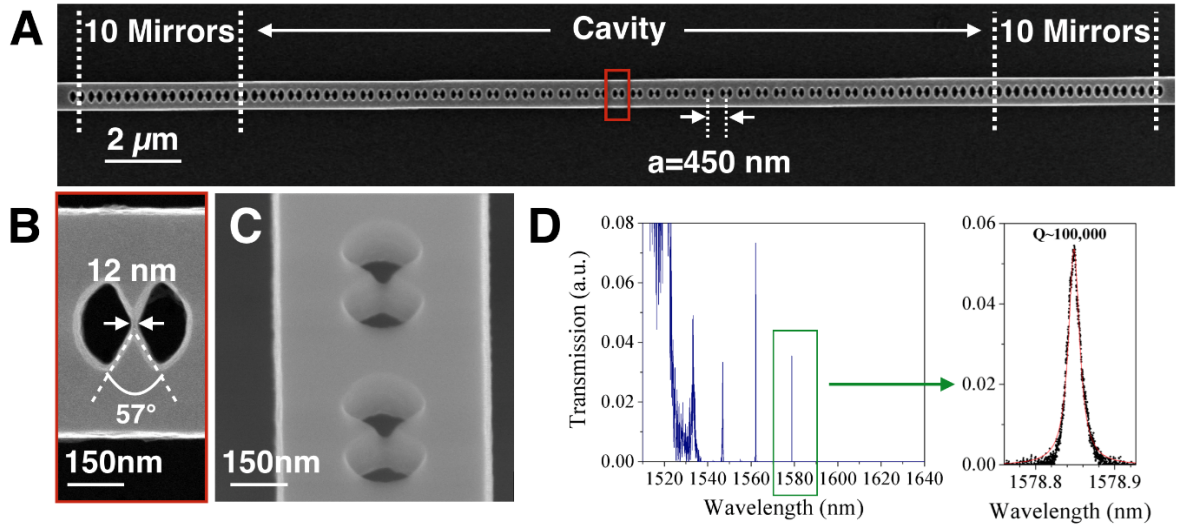

**Fig. 3. Transmission of fabricated bowtie photonic crystal.** (**A**) SEM image of the bowtie photonic crystal. (**B**) Zoomed-in image of a single unit cell in the red box in (**A**). (**C**) Tilted SEM image of an undercut bowtie photonic crystal revealing the out-of-plane profile. (**D**) Measured transmission spectrum. The fundamental mode has Q ~ 100,000 at $\lambda$ = 1578.85 nm. The second order and third order peaks are located at 1562.20 nm and 1546.96 nm, with Q-factors of 21,800 and 5,156, respectively.

**Figure 3A** shows a scanning electronic microscope (SEM) image of the fabricated bowtie photonic crystal with 20 tapering unit cells and 10 mirror unit cells on each side of the central cavity unit cell. The width of the bowtie tip connection is estimated to be approximately 12 nm (**Fig. 3B**). Given that there are only a few pixels comprising the bowtie tip, this value has a relatively large error bar of ±5 nm. The bowtie angle is estimated to be approximately 57°. The radii of center and mirror unit cells are measured to be 147 nm and 190 nm. The width of waveguide is measured to be 691 nm. **Figure 3C** shows a titled SEM image that clearly reveals the v-shaped structure at the bowtie tip; the unit cells shown in this image are from a photonic crystal fabricated by the same process as the one in **Fig. 3A** but released from the oxide substrate using a buffered hydrofluoric acid etch. Transmission measurements carried out on the bowtie photonic crystal show that the fundamental mode at $\lambda_r$ = 1578.85 nm has a loaded Q of approximately $1\times10^5$ (**Fig. 3D**). The modes supported by the photonic crystal are located near the short wavelength band edge (~ 1520 nm), which is consistent with design (**Fig. 3D and Fig. 2B**). The transmission intensity of the resonance peaks is low compared to that of the band edge (**fig. S3**) due to the high mirror confinement in the cavity. We anticipate that higher Q resonances can be designed and measured by using alternate coupling techniques that allow light to be coupled directly into the cavity instead of first passing through the mirror unit cells (e.g., side coupling or evanescent coupling from a fiber).

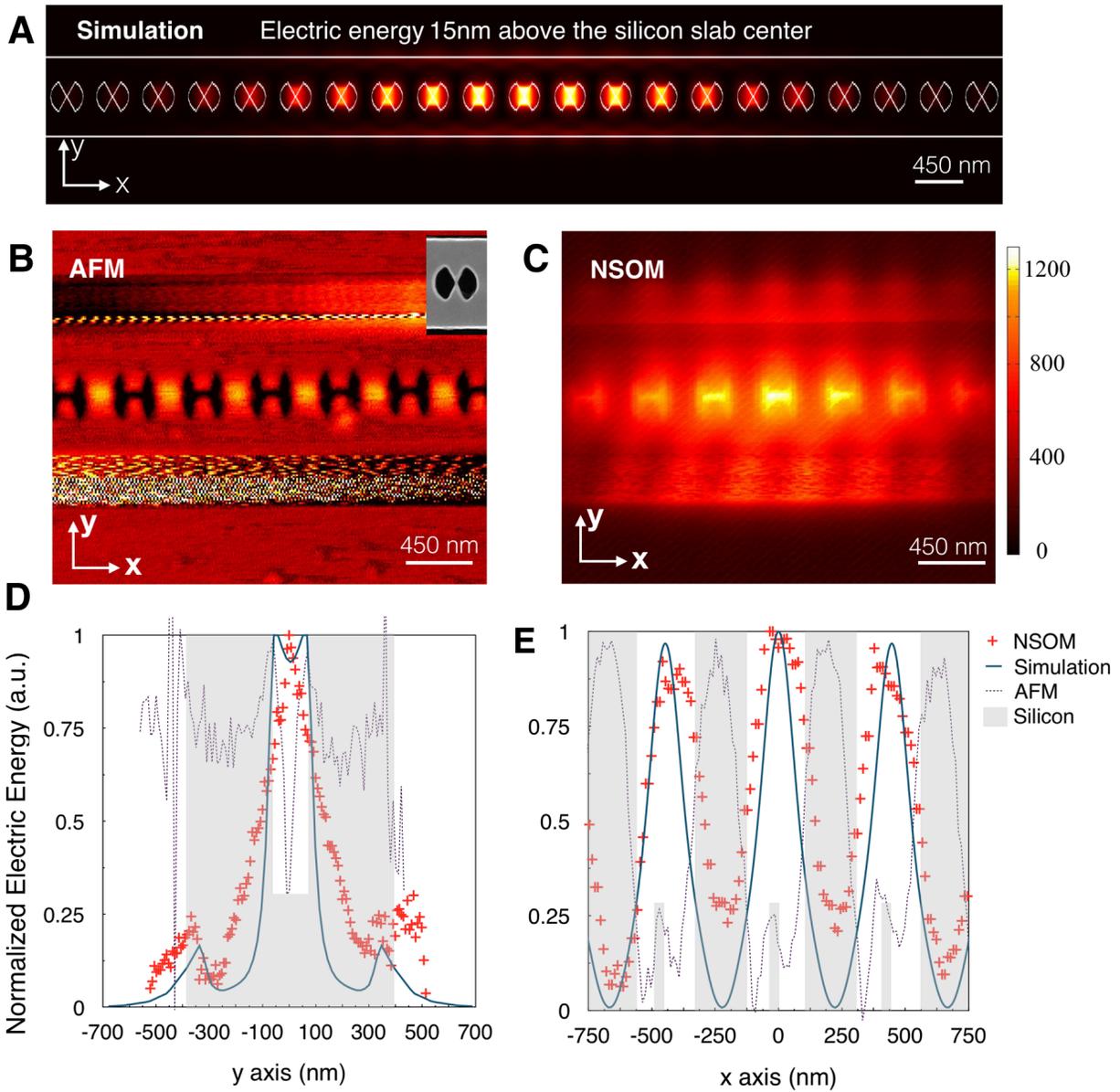

**Fig. 4. Analysis of spatial confinement via NSOM measurements.** (**A**) Schematic of bowtie photonic crystal cavity with overlay of simulated electric energy 15 nm above the silicon surface where the NSOM measures the scattered field. (**B**) AFM measurement and (**C**) corresponding electric energy distribution as measured by NSOM. The inset in (B) shows a higher resolution SEM image of one of the bowtie unit cells from the measured cavity. (**D** - **E**) Simulated and

NSOM-measured near-field profile along *y* direction and *x* direction, respectively, along with superimposed AFM line scan.

To experimentally verify the simulated optical field distribution of the bowtie photonic crystal resonators, we used apertureless near-field scanning optical microscopy (NSOM) to probe the resonance mode (27). Considering the practical constraints of the NSOM system, the structure shown in **Fig. 3** is not ideal for NSOM measurements due to the narrow resonance linewidth (~ 15 pm) and low transmission intensity (~ 0.04). Accordingly, a bowtie photonic crystal is designed and fabricated with reduced mirror confinement (5 mirror unit cells on each side of the cavity) to increase both the linewidth (~ 50 pm) and transmission intensity (~ 0.2), as shown in **fig. S4**. Because the mode is tightly confined in the cavity, the mode volume is not changed by reducing the number of mirror unit cells ($V_m = 6.09 \times 10^{-4}$ µm³ for 5 mirror unit cells). Consequently, conclusions drawn from NSOM measurements on the 5 mirror unit cell bowtie photonic crystal are also applicable to the higher Q cavities with additional mirror unit cells. The NSOM operates in a tapping mode in which the tip oscillates from 0 to 30 nm above the sample surface. Therefore, the near-field measured by the NSOM does not directly correspond to the calculated mode volume within the bowtie (**Fig. 2E**). In order to correlate experiment and simulation, we simulate the electric energy localization in the central unit cell of bowtie photonic crystal at 15 nm above the silicon surface as an estimate of the expected average scattering field that can be detected by the NSOM (**Fig. 4A**). The calculated size of the electric energy localization, which is estimated by the full-width-at-half-maximum (FWHM) of the electric energy distribution in the unit cell, linearly increases from the center of the silicon slab (**fig. S5**) and is estimated to be ~ 183 nm along the *y* direction and 143 nm along the *x* direction at a distance of 15 nm above the silicon surface. **Figure 4B** and **C** show the atomic force microscope (AFM) topology and simultaneously measured optical near field mapping of the bowtie photonic crystal cavity, respectively. The shape of the bowties shown in **Fig. 4B** are distorted due to the resolution limit of the AFM mode of the NSOM when measuring bowtie features that resides below the surface of the sample using a tapping mode above the surface of the sample. The SEM image shown in the inset of **Fig. 4B** reveals the actual shape of one of the bowtie unit cells in this sample. The measured electric energy (**Fig. 4C**) is confined at the bowtie tips in agreement with the simulated electric energy (**Fig. 4A**) distribution in the cavity. **Figure 4D** and **E** show the measured electric energy profiles along *x*- and *y*-slices of the central unit cell. We identify the silicon region as the shaded area in **Fig. 4D** and **E**, based on AFM measurements (dotted line in **Fig. 4, D and E**). The simulated electric energy profiles are shown by the blue curves in **Fig. 4D** and **E**. The NSOM measured profile (red markers) along the *y*-slice through the center of the cavity shows a sharp tip of the field at the bowtie center, indicating a concentrated electric energy. The size is estimated to be ~ 175 nm by considering the FWHM of the electric energy distribution in the central cavity unit cell, which is in agreement with simulation. The NSOM measured field profile along the *x*-slice through the center of the cavity has a FWHM of ~ 267 nm. We attribute the discrepancy between the experimental and simulation results to a combination of the influence of the NSOM tip itself on the electric energy distribution and the fact that NSOM preferably detects the $E_z$ signal, while the photonic crystal is designed for TE polarized light (i.e., mainly $E_y$ component). Importantly, we note that the measured dimensions of the electric energy localization along the *x*- and *y*-directions of the bowtie photonic crystal are within a factor of two of those of plasmonic resonators measured using NSOM systems, as shown in Table S1 (*14, 28, 29*). Moreover, although the field distribution for the bowtie photonic crystal and plasmonic bowties is different, the calculated mode volume, which in the case of the photonic crystal spans multiple unit cells, is nearly identical.

Our work demonstrates that a dielectric resonator can serve as a promising alternative to lossy metals for extreme light concentration and manipulation. Further optimization of the design and fabrication parameters may lead to bowtie photonic crystals with an even higher $Q/V_m$ ratio. We believe our work provides two specific contributions for future research. First, it showcases the power of combining subwavelength dielectric structure with photonic band theory. Engineering the photonic band gap overcomes the challenges encountered while confining optical modes in nano-scatterers, enabling the design of nano-scatterers that are able to precisely adjust the optical mode distributions at subwavelength scales. Second, we prove it is possible in a single, low-loss structure to achieve a mode volume commensurate with plasmonic elements while maintaining a quality factor that is characteristic of traditional photonic crystal cavities. Such an unprecedented strong light-matter interaction platform can facilitate the advancement of science in a broad range of applications.

**References and Notes:**

1. S. Noda, M. Fujita, T. Asano, Spontaneous-emission control by photonic crystals and nanocavities. *Nat Photonics* **1**, 449–458 (2007).
2. R. F. Oulton *et al.*, Plasmon lasers at deep subwavelength scale. *Nature* **461**, 629–632 (2009).
3. H. Altug, D. Englund, J. Vuckovic, Ultrafast photonic crystal nanocavity laser. *Nat Phys* **2**, 484–488 (2006).
4. S. Matsuo *et al.*, High-speed ultracompact buried heterostructure photonic-crystal laser with 13 fJ of energy consumed per bit transmitted. *Nat Photonics* **4**, 648–654 (2010).
5. H. A. Atwater, A. Polman, Plasmonics for improved photovoltaic devices. *Nat Mater* **9**, 205–213 (2010).
6. J. A. Schuller *et al.*, Plasmonics for extreme light concentration and manipulation. *Nat Mater* **9**, 193–204 (2010).
7. M. A. Green, S. Pillai, Harnessing plasmonics for solar cells. *Nat Photonics* **6**, 130–132 (2012).

8. M. L. Brongersma, Y. Cui, S. H. Fan, Light management for photovoltaics using high-index nanostructures. *Nat Mater* **13**, 451–460 (2014).

9. G. T. Reed, G. Mashanovich, F. Y. Gardes, D. J. Thomson, Silicon optical modulators. *Nat Photonics* **4**, 518−526 (2010).

10. Q. F. Xu, B. Schmidt, S. Pradhan, M. Lipson, Micrometre-scale silicon electro-optic modulator. *Nature* **435**, 325−327 (2005).

11. J. C. Rosenberg *et al.*, A 25 Gbps silicon microring modulator based on an interleaved junction. *Opt Express* **20**, 26411−26423 (2012).

12. A. Melikyan *et al.*, High-speed plasmonic phase modulators. *Nat Photonics* **8**, 229–233 (2014).

13. K. Nozaki *et al.*, Sub-femtojoule all-optical switching using a photonic-crystal nanocavity. *Nat Photonics* **4**, 477–483 (2010).

14. Y. Luo *et al.*, On-Chip hybrid photonic–plasmonic light concentrator for nanofocusing in an integrated silicon photonics platform. *Nano Lett* **15**, 849–856 (2015).

15. M. K. Kim *et al.*, Squeezing photons into a point-like space. *Nano Lett* **15**, 4102–4107 (2015).

16. J. C. Ndukaife, V. M. Shalaev, A. Boltasseva, Plasmonics—turning loss into gain. *Science* **351**, 334–335 (2016).

17. J. Yan *et al.*, Directional Fano resonance in a silicon nanosphere dimer. *ACS Nano* **9**, 2968–2980 (2015).

18. R. M. Bakker *et al.*, Magnetic and electric hotspots with silicon nanodimers. *Nano Lett* **15**, 2137–2142 (2015).

19. R. Regmi *et al.*, All-dielectric silicon nanogap antennas to enhance the fluorescence of single molecules. *Nano Lett* **16**, 5143−5151 (2016).

20. A. F. Koenderink, A. Alù, A. Polman, Nanophotonics: shrinking light-based technology. *Science* **348**, 516−521 (2015).

21. E. Kuramochi *et al.*, Ultrahigh-Q one-dimensional photonic crystal nanocavities with modulated mode-gap barriers on SiO2 claddings and on air claddings. *Opt Express* **18**, 15859−15869 (2010).

22. Q. Quan, P. B. Deotare, M. Loncar, Photonic crystal nanobeam cavity strongly coupled to the feeding waveguide. *Appl Phys Lett* **96**, 203102 (2010).

23. Q. M. Quan, M. Loncar, Deterministic design of wavelength scale, ultra-high Q photonic crystal nanobeam cavities. *Opt Express* **19**, 18529–18542 (2011).

24. P. Seidler, K. Lister, U. Drechsler, J. Hofrichter, T. Stoferle, Slotted photonic crystal nanobeam cavity with an ultrahigh quality factor-to-mode volume ratio. *Opt Express* **21**, 32468–32483 (2013).

25. J. D. Ryckman, S. M. Weiss, Low mode volume slotted photonic crystal single nanobeam cavity. *Appl Phys Lett* **101**, 071104 (2012).

26. S. Hu, S. M. Weiss, Design of photonic crystal cavities for extreme light concentration. *ACS Photonics* **3**, 1647–1653 (2016).

27. Materials and methods are available as supplementary materials on Science website.

28. Apuzzo, A., et al., Observation of near-field dipolar interactions involved in a metal nanoparticle chain waveguide. *Nano Lett* **13**, 1000–1006 (2013).

29. V. A. Zenin *et al.*, Boosting local field enhancement by on-chip nanofocusing and impedance-matched plasmonic antennas. *Nano Lett* **15**, 8148–8154 (2015).

**Acknowledgments:**

This work was supported in part by the National Science Foundation (ECCS1407777). The authors thank G. Gaur, K. J. Miller, C. Xiong, B. Peng and J. S. Orcutt for helpful discussions. The photonic crystal bowtie cavities were fabricated in the Microelectronics Research Laboratory (MRL) at the IBM T. J. Watson Research Center. The authors are grateful to the MRL staff for their contributions to the success of this work. Simulations presented in this work were conducted in part using the resources of the Advanced Computing Center for Research and Education at Vanderbilt University, Nashville, TN. Additional computations were carried out with the resources of the HPC Center of Champagne-Ardenne ROMEO. Equipment and technical support in the

**Supplementary Materials:**

Materials and Methods

Fig. S1 to S5

Table S1